\documentclass[preprint,eqsecnum,aps]{revtex4}
\usepackage{epsfig}
\usepackage{graphicx}%
\usepackage{dcolumn}
\usepackage{amsmath}

\makeatletter
\def\btt#1{\texttt{\@backslashchar#1}}%
\DeclareRobustCommand\bblash{\btt{\@backslashchar}}%
\makeatother

\begin{document}

\title{Phantom-like GCG and the constraints of its parameters via cosmological dynamics}

\author{Jiangang
Hao}\affiliation{Shanghai United Center for Astrophysics (SUCA),
Shanghai Normal University, 100 Guilin Road, Shanghai 200234,
China \\ Department of Physics, the University of Michigan, 500 E.
University Ave. Ann Arbor, MI 48109, USA}
\author{Xin-zhou Li}\email{kychz@shnu.edu.cn}
\affiliation{Shanghai United Center for Astrophysics (SUCA),
Shanghai Normal University, 100 Guilin Road, Shanghai 200234,
China\\
Division of Astrophysics, E-institute of Shanghai Universities,
Shanghai Normal University, 100 Guilin Road, Shanghai 200234,
China
}%

\date{\today}

\begin{abstract}
We extend the equation of state of GCG (Generalized Chaplygin Gas)
to $w<-1$ regime and show, from the point of view of dynamics,
that the parameters of GCG should be in the range of $0<\alpha<1$.
Also, dynamical analysis indicate that the phase $w_{g}=-1$ is a
dynamical attractor and the equation of state of GCG approaches it
from either $w_{g}>-1$ or $w_{g}<-1$ depending on the choice of
its initial cosmic density parameter and the ratio of pressure to
critical energy density.
\end{abstract}


\maketitle

\section*{1. Introduction}

CMB anisotropy \cite{bennett,Netterfield,Halverson}, Supernovae
\cite{riess, perlmutter,tonry} and SDSS\cite{SDSS} strongly
indicate that our universe is spatially flat, with two thirds of
the energy contents resulted from dark energy, a substance with
negative pressure and can make the universe expanding in an
accelerating fashion. Candidates for dark energy have been
proposed as vacuum energy, quintessence
\cite{ratra,Wetterich,Frieman,Ferreira,coble,Steinhardt,Peebles},
phantom \cite{caldwell} and GCG\cite{GCG} which is stemmed from
the Chaplygian gas\cite{chgas}. Present observation data constrain
the the range of the equation of state of dark energy as
$-1.38<w<-0.82$ \cite{melchiorri}, which indicates the possibility
of dark energy with $w<-1$, debuted as Phantom\cite{caldwell}. The
realization of $w<-1$ could not be achieved by scalar field with
positive kinetic energy and thus the negative kinetic energy is
introduced although it violates some well known energy conditions
\cite{carroll}. Another important consequence of Phantom is the
Big rip \cite{bigrip} or Big smash \cite{McInnes} phase, in which
the scale factor of the Universe goes to infinity at a finite
cosmological time. The cosmological implications of Phantom have
been widely
studied\cite{sahni,schulz,haonew,maor,Frampton,Gibbons,hao1,Sami,
Feinstein,Chimento,Dabrowski,Nojiri,Johri,Cline,Lu,Gonzalez-Diaz,Stefancic,brevik}
and the Phantom model with Born-Infeld type Lagrangian has been
proposed \cite{hao3} and its generalization to brane world has
been done in Ref. \cite{liu}.

GCG has a very simple equation of state,
$p_{g}=-\frac{M^{4(\alpha+1)}}{\rho_{g}^{\alpha}}$, which yields
an analytically solvable cosmological dynamics if Universe is GCG
dominated. Another ambition of introducing GCG is to unify unify
dark energy and dark matter into one equation of state, also known
as quartessence\cite{malker}. However, detailed numerical analysis
turns out to disfavor the dark matter modelled by the GCG equation
of state\cite{sandvik}. But no observation so far rule out the
possibility of GCG as dark energy. It is quite possible for our
Universe to contain a dark energy component modelled by GCG as
well as another baryotropic fluid component mimicked by the
equation of state $p_{\gamma}=\gamma \rho_{\gamma}$. Although the
cosmological GCG system is analytically solvable when GCG is
dominant, it is no longer possible when another component is
present. We therefore resort to the phase analysis with which one
can gain many important information without solving the dynamical
equations.

Previous study of GCG are focus on the case that $w_{g}>-1$. In
fact, if we accept the notion that dark energy is modelled by a
perfect fluid, then, current observations do not exclude the
possibility of $w_{g}<-1$, instead, they even favor
it\cite{melchiorri}. Thus, in the parameter space of the GCG
model, one should not exclude the regime $w_{g}<-1$ although this
range of equation of state can not be smoothly continued from the
$w_{g}>-1$ regime. In this letter, we generalize the idea of GCG
by considering the $w_{g}<-1$ regime of its equation of state. We
show that the system could reach the late time de Sitter attractor
from either $w_{g}>-1$ or $w_{g}<-1$ depending on the choice of
the initial energy density and pressure. When $w_{g}<-1$, it will
behave as phantom with a late time de Sitter attractor, and
therefore won't lead to the catastrophic big rip.

The equation of state of GCG has two free parameters, $\alpha$ and
$M$, which could be fixed, in principle, by fitting the model to
the Supernovae or CMB data\cite{GCG}. However, when dealing with
these fitting, we need to narrow down the possibilities of the
range of the parameters to facilitate the numerical analysis. This
is just part of the purpose of this current work, in which we
constrain the parameter $\alpha$ from the cosmological dynamics of
the system as well as the requirement of sound speed. Our result
is in agreement with the numerical results obtained by other
authors.

\section*{2. Autonomous system}

A general study the phase space system of phantom scalar field in
FRW universe has been given in Ref.\cite{haonew}. For the GCG
cosmological dynamical system, the corresponding equations of
motion and Einstein equations could be written as,

\begin{eqnarray}\label{sys}
&&\dot{H}=-\frac{\kappa^2}{2}(\rho_\gamma+p_\gamma+\rho_{g}+p_{g})\nonumber\\
&&\dot{\rho_\gamma}=-3H(\rho_\gamma+p_\gamma)\nonumber\\
&&\dot{\rho_{g}}=-3H(\rho_{g}+p_{g})\nonumber\\
&&H^2=\frac{\kappa^2}{3}(\rho_\gamma+\rho_{g})
\end{eqnarray}

\noindent where $\kappa^2=8\pi G$, $\rho_{\gamma}$ is the density
of fluid with a baryotropic equation of state
$p_{\gamma}=(\gamma-1)\rho_{\gamma}$, where $0\leq \gamma\leq2$ is
a constant that relates to the equation of state by
$w_{\gamma}=\gamma-1$; $\rho_{g}$ is the energy density of GCG
with an equation of state

\begin{equation}\label{eosquart}
p_{g}=-\frac{M^{4(\alpha+1)}}{\rho_{g}^{\alpha}}
\end{equation}

\noindent The over dot represents derivative with respect to
cosmic time $t$, and $H$ is Hubble parameter.

To analyze the dynamical system, we rewrite the equations with the
following dimensionless variables

\begin{eqnarray}\label{vari}
x&=&\frac{\kappa^2\rho_{g}}{3H^2}\nonumber\\
y&=&\frac{\kappa^2p_{g}}{3H^2}\nonumber\\
N&=&\ln a
\end{eqnarray}

\noindent The dynamical system will then reduce to

\begin{eqnarray}\label{auto}
\frac{dx}{dN}&=&-3(x+y)+3x[\gamma(1-x)+x+y]\nonumber\\
\frac{dy}{dN}&=&3\alpha(y+y^2/x)+3y[\gamma(1-x)+x+y]
\end{eqnarray}

\noindent Accordingly, the Friedman equation yields

\begin{equation}\label{constraint}
\Omega_{g}+\Omega_{\gamma}=1
\end{equation}

\noindent where $\Omega_{g}\equiv x$ and $\Omega_{\gamma}\equiv
\frac{\kappa^2\rho_{\gamma}}{3H^2}$ are the cosmic density
parameters for GCG and baryotropic fluid respectively. The
equation of state for the scalar fields could be expressed in
terms of the new variables as

\begin{equation}\label{equaofstate}
 w_{g}=\frac{p_{g}}{\rho_{g}}=\frac{y}{x}
\end{equation}

\noindent and the sound speed is

\begin{eqnarray}\label{sound}
c_s^2= -\alpha\frac{y}{x}
\end{eqnarray}

The critical points of the system are $(x, y)=(1, 0)$ and $(1,
-1)$ which correspond to a matter dominated phase and vacuum
energy dominated phase respectively. If we linearize the system
near its critical points and then translate the system to origin,
we could readily write the first order perturbation equation as

\begin{eqnarray}\label{perturb}
\textbf{U}'=\emph{A}\cdot \textbf{U}
\end{eqnarray}

\noindent where $\textbf{U}$ is a 2-column vector consist of the
perturbations of $x$ and $y$. $A$ is a $2\times2$ matrix
\begin{widetext}
\begin{eqnarray}
A=\left(%
\begin{array}{cc}
  -3+3\gamma-6\gamma x+6x+3y &  -3+3x  \\
   -\frac{3\alpha y^2}{x^2}-3\gamma y+3y &  3\alpha+\frac{6\alpha y}{x}+3\gamma-3\gamma x+3x+6y  \\
\end{array}%
\right)
\end{eqnarray}
\end{widetext}

\noindent The stability of the critical points is determined by
the eigenvalues of the matrix $A$ at the critical points. For the
point $(1, 0)$, the two eigenvalues are

\begin{eqnarray}\label{eigen1}
\lambda_{1}&=&3-3\gamma\\\nonumber \lambda_{2}&=& 3\alpha
\end{eqnarray}

\noindent So, it may be stable if $\alpha<0$ and $\gamma\geq 1$.
However, since we want the GCG behaves as dark energy, it won't be
appropriate if $(1, 0)$ corresponds to a stable attractor phase.
In other words,
 $\alpha<0$ should not be considered in the real models. While for
the critical point $(1, -1)$, the corresponding eigenvalues of
matrix $A$ are

\begin{eqnarray}\label{eigen2}
\lambda_{1}&=&-3(1+\alpha)\\\nonumber \lambda_{2}&=&-3\gamma
\end{eqnarray}

\noindent It is clear that $(1, -1)$ is stable for $\alpha>-1$.
This critical point corresponds to a phase that GCG is dominant
($\Omega_{g}=1$) and its equation of state is $w=-1$. So, it is a
late time de Sitter attractor\cite{hao2}. Combine the above two
constraints as well as the requirements of $c_s^2<1$ at the de
Sitter attractor, we can readily reach the constraint for the
$\alpha$ parameter should be $0<\alpha<1$.

In the following, we study the above dynamical system numerically.
For definite, we choose the parameters as $\gamma=1$ and
$\alpha=0.5$. The initial $x$ and $y$ are chosen as shown in Table
I and the results are contained in the following Fig.1-Fig.4. From
Fig.2, one can observe that for different initial $\rho_{g}$ and
$p_{g}$, the equation of state $w_{g}$ could approach to the de
Sitter phase ($w_{g}=-1$) from either $w_{g}>-1$ or $w_{g}<-1$.
The critical case will be that the initial $x$ and $y$ are such
chosen that $w_{g}=y/x=-1$. In other words, the initial choice of
$w_{g}>-1$ or $<-1$ determines whether the equation of state of
GCG will mimic that of quintessence or phantom. It is worth noting
that if we choose the parameters so that the GCG behaves as
phantom, it is no longer possible to make it behave as matter at
early epoch because the equation of state could not smoothly
evolves from 0 to $w_{g}<-1$. However, in our setup of this paper,
we have included a baryotropic fluid that could be used to mimic
the matter sector of our universe and thus GCG can be considered
only as dark energy.

\begin{center}
\begin{table}
\begin{tabular}{ c  c c c c c c c }
  \hline
  x & 0.14 & 0.15 & 0.16 & 0.17 & 0.18 & 0.19 & 0.20 \\
  \hline
  y & -0.19 & -0.18 & -0.17 & -0.16 & -0.15 & -0.14 & -0.13 \\
  \hline
\end{tabular}
\caption{The initial values of $x$ and $y$ in the plots
Fig.1-Fig.4}
\end{table}
\end{center}

\begin{figure}
\epsfig{file=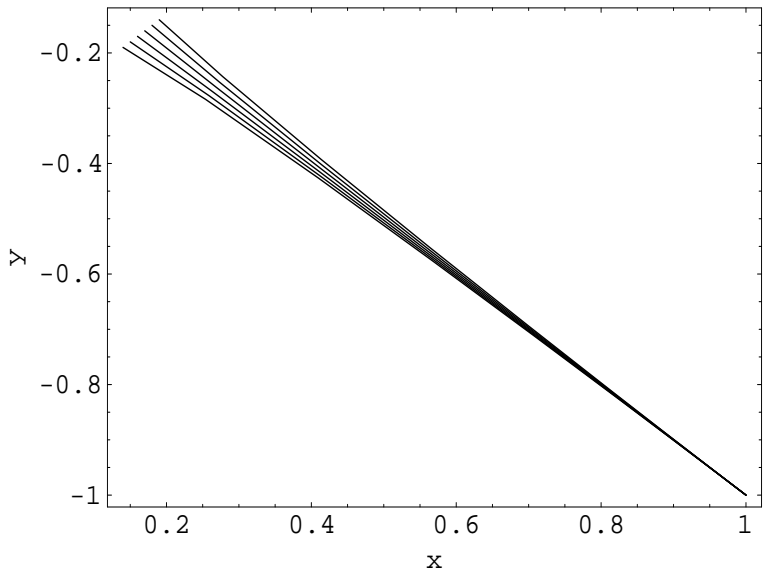,height=3in,width=4in} \caption{The phase
diagram of the GCG system in terms of $x$ and $y$ for different
initial $x$ and $y$.}
\end{figure}

\begin{figure}
\epsfig{file=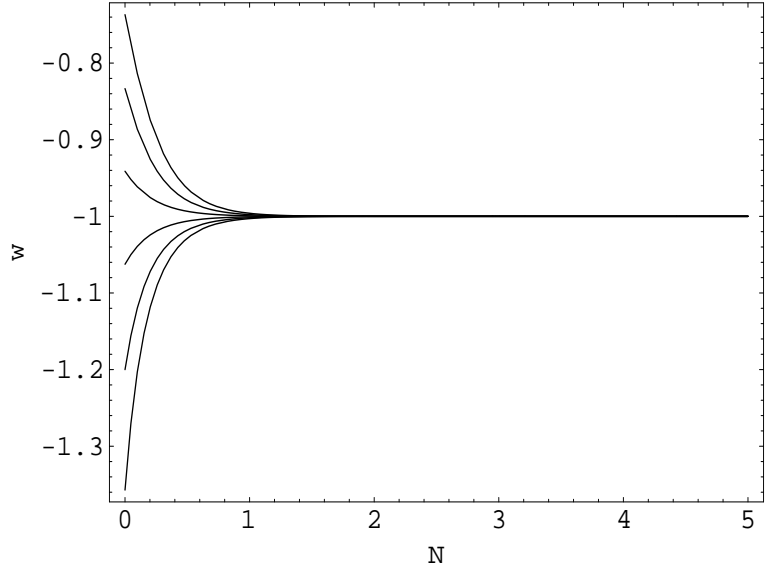,height=3in,width=4in} \caption{The evolution
of the equation of state of GCG for different initial $x$ and $y$.
The curves from bottom to top correspond to the initial conditions
specified in Table I from left to right respectively.}
\end{figure}

\begin{figure}
\epsfig{file=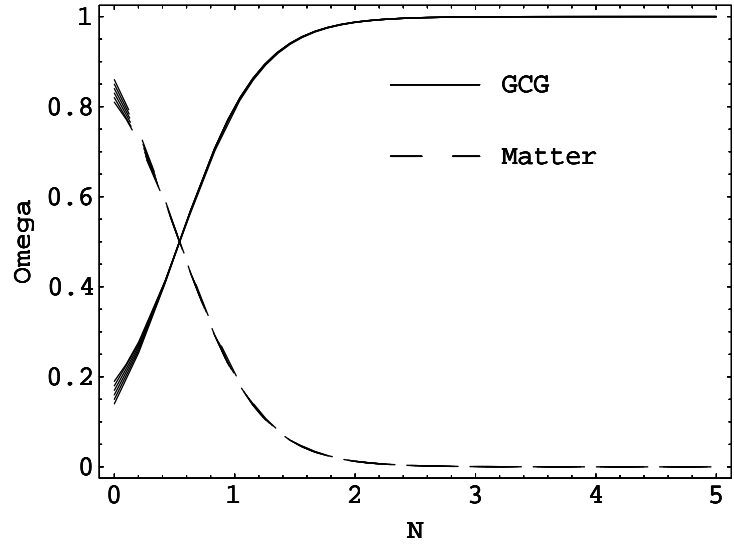,height=3in,width=4in} \caption{The evolution
of the cosmic density parameter for matter $\Omega_{\gamma}$ and
GCG $\Omega_{g}$ respectively at different initial $x$ and $y$.
The plot indicates that the evolution of $\Omega_{g}$ and
$\Omega_{\gamma}$ is not very sensitive to the initial condition
due to the attractor property.}
\end{figure}
\clearpage
\begin{figure}
\epsfig{file=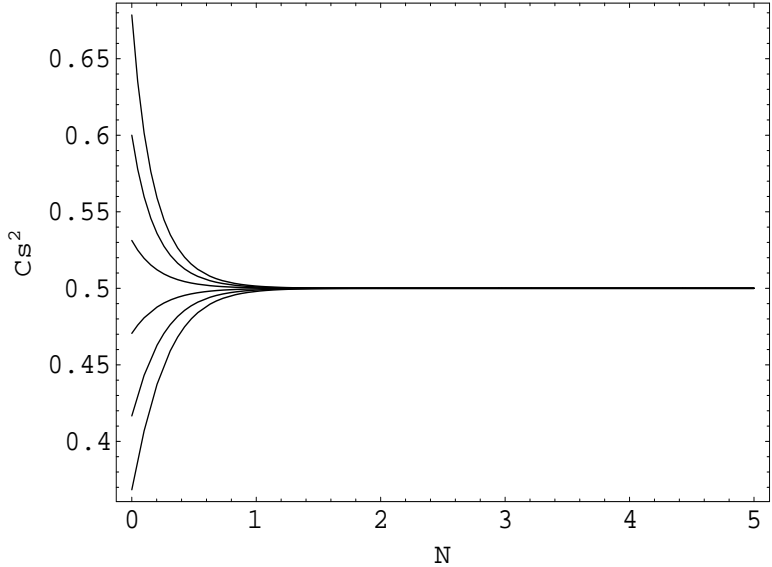,height=3in,width=4in} \caption{The evolution
of sound speed of GCG $c_s^2$ for different initial $x$ and $y$.
The curves from top to bottom correspond to the initial conditions
specified in Table I from left to right respectively.}
\end{figure}

\section*{3. Conclusion and Discussion}

In this letter, we analyze the dynamical evolution of GCG for
different parameters and initial conditions. We show that
different initial $x$ and $y$ will lead to different tracks
($w_{g}>-1$ and $w_{g}<-1$) for the equation of state $w_{g}$ to
approach the de Sitter attractor phase ($w=-1$). That is to say,
the GCG could mimic both quintessence and phantom during its
evolution depending on the initial conditions. We also give
constraint to the parameter of the model as $0<\alpha<1$ from the
requirement of its dynamics and sound speed.

On the other hand, the existing studies of GCG and its fitting to
Supernovae data focus on $w_{g}>-1$\cite{malker,macorra}. But in
fact, observations do not exclude the possibility of $w_{g}<-1$.
So the GCG with $w_{g}<-1$ should also be considered and its
fitting to SNeIa data will be carried out in a preparing work.

\vspace{0.8cm} \noindent ACKNOWLEDGEMENT: This work is supported
by National Nature Science Foundation of China under Grant No.
10473007.

\end{document}